\documentclass[twocolumn,aps,superscriptaddress,floatfix]{revtex4}
\usepackage{amsmath}
\usepackage{graphicx}
\usepackage{dcolumn}
\usepackage{bm}
\usepackage{amssymb}
\usepackage{color}
\usepackage{xcolor}
\usepackage{xfrac}

\begin{document}

\title{Total cost of operating an information engine}

\author{Jaegon Um}
\affiliation{Universit\"at W\"urzburg, Fakult\"at f\"ur  Physik und Astronomie, 97074  W\"urzburg, Germany,}
\affiliation{Quantum Universe Center, Korea Institute for Advanced Study, Seoul 130-722, Korea}
\author{Haye Hinrichsen}
\affiliation{Universit\"at W\"urzburg, Fakult\"at f\"ur  Physik und Astronomie, 97074  W\"urzburg, Germany,}
\author{Chulan Kwon}
\affiliation{Department of Physics, Myongji University, Yongin 449-728, Korea}
\author{Hyunggyu Park}
\affiliation{School of Physics, Korea Institute for Advanced Study,
Seoul 130-722, Korea}

\def\d{{\rm d}}
\def\0{\emptyset}
\def\pi{p^{(i)}}
\def\pf{p^{(\infty)}}

\begin{abstract}
We study a two-level system controlled in a discrete feedback loop, modeling both the system and the controller in terms of stochastic Markov processes. We find that the extracted work, which is known to be bounded from above by the mutual information acquired during measurement, has to be compensated by an additional energy supply during the measurement process itself, which is bounded by the same mutual information from below. Our results confirm that the total cost of operating an information engine is in full agreement with the conventional second law of thermodynamics. We also consider the efficiency of the information engine as function of the cycle time and discuss the operating condition for maximal power generation. Moreover, we find that the entropy production of our information engine is maximal for maximal efficiency, in sharp contrast to conventional reversible heat engines.
\end{abstract}

\maketitle
\parskip 1mm

\section{Introduction}

In 1867, J.C. Maxwell created a thought experiment to demonstrate a possible violation of the second law of thermodynamics: A thermally isolated container with a gas is divided into two parts and a fictitious demon opens or closes a door between the two parts depending on the velocity of approaching particles, creating an increase of the temperature in one of the compartments~\cite{HistoricalReview}.

Smoluchowski was the first to provide an explanation why Maxwell's demon does not work. To this end, he modeled the demon mechanically by a trapdoor combined with a gentle spring. The trapdoor acts as a valve in such a way that fast particles coming from one side can open the door while slow ones cannot, leading to a pressure difference. However, taking the full dynamics of the apparatus into account, Smoluchowski could demonstrate that the energy of the spring system itself equilibrates at such a high energy that it opens and closes essentially randomly, leading to the same pressure difference as if the trapdoor was always open.

\begin{figure}
\centering\includegraphics[width=75mm]{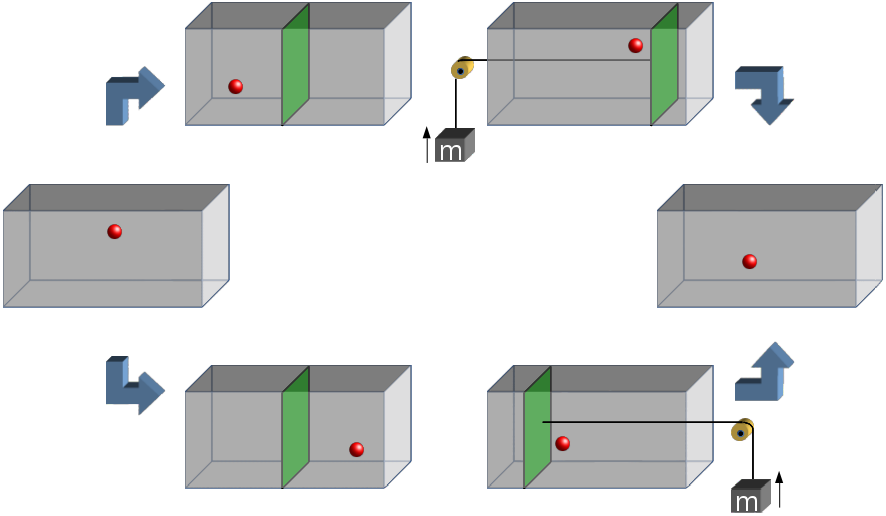}
\vspace{-2mm}
\caption{Szil\'ard engine. In a box with a single molecule, a piston is inserted in the middle. Depending on the location of the particle, the piston is pushed in one direction, allowing us to extract work. However, placing the load correctly requires 1 bit of information about the position of the particle. At the end, the piston is removed at zero cost.}
\label{fig:szilard}
\end{figure}

In 1928, L. Szil\'ard refined the concept of Maxwell's demon, suggesting what is known today as the \textit{Szil\'ard engine}~\cite{Szilard}. Starting point is a box that contains only one particle (see Fig.~\ref{fig:szilard}). If a wall is inserted in the middle, the particle will be in one of the two parts. Expanding the volume isothermally to its original size by moving the shutter into the empty half of the box one can extract the work $ W=k_B T \ln 2$. However, this requires to know in which compartment the particle actually is, demonstrating that the possession of information can be converted into physical work. Thus, in order to keep a Szil\'ard engine running, a closed loop of measurement and feedback is needed. Very recently, such a feedback scheme could be realized experimentally for the first time~\cite{SzilardExperiment}.

Since the feedback mechanism in the Szil\'ard engine was perceived as information processing at zero cost, it seemed to produce usable work from nothing, violating the second law.  However, as shown by Landauer~\cite{Landauer} in 1961, any irreversible logical operation requires to apply a well-defined minimum of work. In case of the Szil\'ard engine, the demon has to store the information about the particle position in a single bit. According to Landauer's principle, resetting this bit requires to convert a work of $k_B T \ln 2$ into heat. As shown by Bennett~\cite{Bennett}, who realized measurement and feedback as reversible processes, this extra work for resetting restores the second law.

Recently Maxwell's demon attracted renewed attention as it was shown that a system in a feedback loop obeys an integral fluctuation theorem (IFT) of the form
\begin{equation}
\langle e^{\beta  W_{ex} - \Delta I}\rangle=1 \,,
\end{equation}
implying $ \beta \langle  W_{ex} \rangle \leq  \langle\Delta I\rangle$, where $W_{ex}$ is the extracted work during feedback, $\beta$ is the inverse temperature, and $\Delta I$ is the gained mutual information between system and demon during the measurement~\cite{SagawaUeda1,SagawaUeda2,SagawaUeda3}. This fluctuation theorem implies that in each cycle of the engine the extracted work is limited by the gained mutual information, -- a highly plausible result that nicely demonstrates the equivalence of thermodynamic work and information.

Subsequently this remarkable result was made more specific in various ways. For example, it was shown that one can construct feedback schemes that satisfy the inequality sharply~\cite{SagawaUeda4,Horowitz}. Moreover, the IFT was generalized to schemes with finite-time relaxation~\cite{Abreu,Bauer} and continuous feedback schemes~\cite{Sandber}. Very recently, the generalized IFT could also be confirmed experimentally~\cite{Koski}.

The arising problem with these generalized Jarzynski equalities is that the tightness of the bound seems to depend on the specific feedback scheme such as discrete and continuous feedback as well as memory tape models. This led Barato \textit{et al.} to look for a unifying master IFT~\cite{Barato1,Barato2,Barato3,Barato4}. To achieve this, they duplicated the configuration space, modeling bit flips of the memory by transitions between the two replicas. Doing so they followed Smoluchowski's original idea of modeling the whole feedback loop as a physical device, defined as a stochastic Markov process.

In this paper, we follow these lines of thought, being interested in the \textit{total} cost of operating an information engine as described above. Instead of duplicating the configuration space, we devise physically realizable stochastic processes of the joint system not only for relaxation, but also for the measurement process. We show that the generalized IFT for the relaxation of system is always accompanied by an \textit{opposite} IFT during the measurement carried out by the demon, restoring the second law for the joint system. This implies that a certain minimal amount of work has to be done on the memory, in accordance with Landauer's principle. The calculation of the total cost enables us to derive the efficiency of the information engine as a function of its cycle time. We also discuss the optimal cycle time and corresponding efficiency, at which the extracted power is maximized.

\section{Definition of a minimal model}

\begin{figure}
\centering\includegraphics[width=85mm]{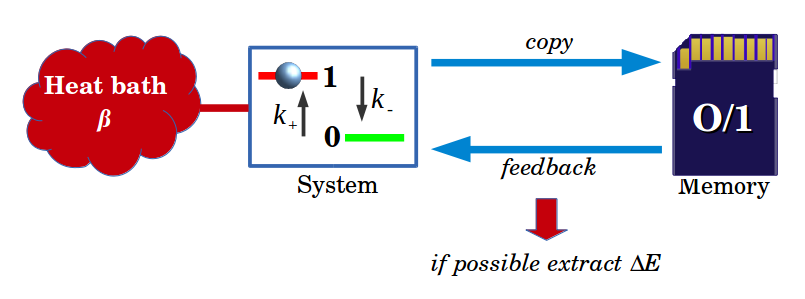}
\vspace{-3mm}
\caption{Schematic drawing of the information engine studied in the present work. The configuration of a thermally relaxing two-state system is measured by a controller and stored in a one-bit memory. If the system is found in the upper level, the controller induces a transition back to the ground state, extracting the energy $\Delta E$. }
\label{fig:engine}
\end{figure}

In what follows we consider a system with two different energy levels separated by $\Delta E$. The system is coupled to a heat bath with inverse temperature $\beta$. The controller (demon) is implemented as a 1-bit memory. We devise a discrete feedback scheme evolving in three steps (see Fig.~\ref{fig:engine}). First the system relaxes thermally without external influence. Then the actual energy level, denoted by $0$ and $1$, is copied to the memory of the controller. Finally, if the memory bit is $1$, the controller induces a flip of the system $1 \to 0$, extracting the energy $\Delta E$, otherwise it does nothing.

Following Barato and Seifert~\cite{Barato4}, we are aiming to model these steps by stochastic Markov processes, including both the system and the memory. This defines a four-dimensional configuration space in which each step can be represented by a simple $4\times 4$ matrix. In the following, we discuss each of the three steps in detail:\\

\begin{center} \textbf{1. Relaxation}\\ \end{center}
During relaxation, the system flips randomly according to the rules
\begin{eqnarray*}
0 \to 1 &\quad& \mbox{with rate } k_+\\
1 \to 0 &\quad& \mbox{with rate } k_-\,\,.
\end{eqnarray*}
For convenience, we express these rates in terms of
\begin{equation}
k=k_++k_-\quad\mbox{and}\qquad q=\frac{k_+}{k}
\end{equation}
with $q <\frac{1}{2}$.
After infinite time, the system eventually reaches an equilibrium state with the stationary probability distribution $P^{1}_{stat}=1-P_{stat}^0=q$. This means that the energy difference between the two levels is given by
\begin{equation}
\label{DeltaE}
\Delta E=\beta^{-1} \ln\frac{\bar q}{q}\, \quad\quad (\bar q = 1-q).
\end{equation}
Let us now describe the relaxation process in the composite configuration space of system and memory. Throughout this paper, we will use a canonical configuration basis ordered by
\begin{equation}
\label{basis}
(\mbox{system state},\, \mbox{memory bit}) = \{00,01,10,11\}.
\end{equation}
Since the memory is inactive during relaxation, the time evolution operator $\mathcal L_R$ for relaxation represented in this basis reads
\begin{eqnarray}
\small
 \mathcal L_R &=&
\underbrace{\left(
\begin{array}{cc}
 k_+ & -k_- \\
 -k_+ & k_- \\
\end{array}
\right)}_{\mbox{System}} \otimes
\underbrace{\left(
\begin{array}{cc}
 1 & 0 \\
 0 & 1 \\
\end{array}
\right)}_{\mbox{Memory}}  \nonumber \\
&=&\left(
\begin{array}{cccc}
 k_+ & 0 & -k_- & 0 \\
 0 & k_+ & 0 & -k_- \\
 -k_+ & 0 & k_- & 0 \\
 0 & -k_+ & 0 & k_- \\
\end{array}
\right)\,.
\end{eqnarray}
After the relaxation time $t_R$, the corresponding transition matrix is given by
\begin{equation}
\mathcal T_R(t_R) \;=\; \exp(-\mathcal L_R t_R)\ .
\end{equation}
In the infinite-time relaxation limit ($t_R\to\infty$), this transition matrix reduces to
\begin{equation}
\label{TRI}
\mathcal T_R \;=\; \lim_{t_R \to \infty} \mathcal T_R (t_R) \;=\; \begin{pmatrix}
\bar q & 0 & \bar q & 0 \\
0 & \bar q & 0 & \bar q \\
q & 0 & q & 0 \\
0 & q & 0 & q
\end{pmatrix}.
\end{equation}
\vspace{3mm}

\begin{center} \textbf{2. Measurement}\\ \end{center}
A perfect measurement would faithfully copy the system state to the memory, i.e., if $m=0,1$ denotes the previous memory state and $s=0,1$ the actual system state, it would simply copy $(s,m) \mapsto (s,s)$. However, it is well-known that such a \textit{perfect} measurement is irreversible, leading to a diverging entropy production~\cite{Barato4, Esposito}. Therefore, one usually considers \textit{imperfect} measurements
\begin{eqnarray*}
(s,m) \mapsto (s,s) \hspace{6mm} &\quad\mbox{with prob.}\quad& \bar\epsilon=1-\epsilon\\
(s,m) \mapsto (s,1-s) &\quad\mbox{with prob.}\quad& \epsilon
\end{eqnarray*}
with a small error probability $0<\epsilon<1/2$. Using the basis (\ref{basis}) this corresponds to the transition matrix
\begin{equation}
\label{TMI}
\mathcal T_M =
\begin{pmatrix}
 \bar \epsilon  & \bar \epsilon  & 0 & 0 \\
 \epsilon  & \epsilon  & 0 & 0 \\
 0 & 0 & \epsilon  & \epsilon  \\
 0 & 0 & \bar \epsilon  & \bar \epsilon  \\
\end{pmatrix} .
\end{equation}
Remarkably, this imperfect measurement process can be implemented by a stochastic Markov process as well, since ${\mathcal T}_{M}^2 =\mathcal T_M$. The corresponding time evolution operator reads
\begin{equation}
\label{LM}
\mathcal L_M = k'
\begin{pmatrix}
 \epsilon  & -\bar \epsilon  & 0 & 0 \\
 -\epsilon  & \bar \epsilon  & 0 & 0 \\
 0 & 0 & \bar \epsilon  & -\epsilon  \\
 0 & 0 & -\bar \epsilon  & \epsilon
\end{pmatrix}
\end{equation}
with a rate $k'$ and it is easy to show that the transition matrix ${\mathcal T_M}$ in Eq.~(\ref{TMI}) is retrieved in the limit of infinite measurement time:
\begin{equation}
\mathcal T_M = \lim_{t_M \to \infty} \mathcal T_M (t_M) = \lim_{t_M \to \infty} \mathcal \exp(-\mathcal L_M t_M)\,.
\end{equation}
Thus, we succeeded to implement the second step as a stochastic Markov process as well.

If the memory is considered as being in contact with some heat bath of inverse temperature $\beta$ during the stochastic measurement process, the time evolution defined above implies that the incorrectly measured state $(s,1-s)$ has a higher energy than the correctly measured state $(s,s)$, and that the corresponding energy difference between the two composite states is given by
\begin{equation}
\Delta E' = {\beta}^{-1} \ln \frac{\bar{\epsilon}}{\epsilon}\ .
\end{equation}

\begin{center} \textbf{3. Feedback}\\ \end{center}
%
The purpose of the feedback is to use the information stored in the memory in order to extract energy from the system. If the preceding measurement was faithful, this would mean to perform the transitions
\begin{eqnarray*}
00 \mapsto 00 && \quad\mbox{without extraction of energy}\\
11 \mapsto 01 && \quad\mbox{extracting the work $ W_{ex}=\Delta E$.}
\end{eqnarray*}
These transitions alone would be again irreversible, causing an infinite entropy production. However, if we add symmetric transitions in the (unlikely) case of erroneous measurements, namely,
\begin{eqnarray*}
10 \mapsto 10 && \quad\mbox{without performing work}\\
01 \mapsto 11 && \quad\mbox{performing work, i.e., $ W_{ex}=-\Delta E$,}
\end{eqnarray*}
we obtain the feedback transition matrix
\begin{equation}
\label{TFI}
\mathcal T_F =
\begin{pmatrix}
 \bf 1 & 0 & 0 & 0 \\
 0 & 0 & 0 & \bf 1 \\
 0 & 0 & \bf 1 & 0 \\
 0 & \bf 1 & 0 & 0
\end{pmatrix}\,.
\end{equation}
Thus the feedback process is carried out in such a way that the system state is flipped ($s \mapsto 1-s$) for $m=1$ while it remains unchanged ($ s \mapsto s$) for $m=0$. It is assumed that the feedback transition occurs instantaneously so that the total time $\tau$ of a complete cycle $\mathcal T_R \to \mathcal T_M \to \mathcal T_F$ is given by $\tau = t_R + t_M $.

Since $\mathcal T_F^2=\mathbf{I}$, the feedback is fully reversible, hence it does not produce entropy in the environment. Moreover, it is easy to see that it simply exchanges the second and the fourth component of a vector, and therefore it does not change the joint entropy of system and memory. However, as will be shown below, it generally changes the entropy of the subsystems. 

Due to its reversible nature, the feedback as defined above cannot be implemented as a stochastic Markov process.  However, we would like to point out that it is even possible to implement the feedback physically so that the entire chain of steps is represented cleanly as a sequence of stochastic processes. This can be done by replacing two subsequent cycles $\mathcal T_R \to \mathcal T_M \to \mathcal T_F \to \mathcal T_R \to \mathcal T_M \to \mathcal T_F \to \mathcal T_R $ equivalently by $\mathcal T_R \to \mathcal T_M \to \mathcal T_F \mathcal T_R  \mathcal T_F \to  \mathcal T_F \mathcal T_M \mathcal T_F \to \mathcal T_R$.
Since $\mathcal T_{\tilde R} \equiv\mathcal T_F \mathcal T_R  \mathcal T_F $ and $\mathcal T_{\tilde M}\equiv  \mathcal T_F \mathcal T_M \mathcal T_F$ satisfy the stochasticity condition ($\mathcal {T}_{\tilde R}^2=\mathcal T_{\tilde R}$, $\mathcal {T}_{\tilde M}^2=\mathcal T_{\tilde M}$), the whole sequence of steps can be implemented by stochastic processes, providing a safe ground for the calculation of entropy production, mutual information, work, and heat. Having verified that this description is fully equivalent, we nevertheless keep the explicit feedback for simplicity in the original form.

Since the rates $k$ and $k'$ simply rescale $t_R$ and $t_M$, we will set
\begin{equation}
k=k' :=1 
\end{equation}
throughout the paper. Thus, apart from $t_R$ and $t_M$, the model is controlled by only two parameters, namely, the relaxation parameter $q$ and the error probability $\epsilon$.

\section{Stationary state}

\begin{figure}
\centering\includegraphics[width=85mm]{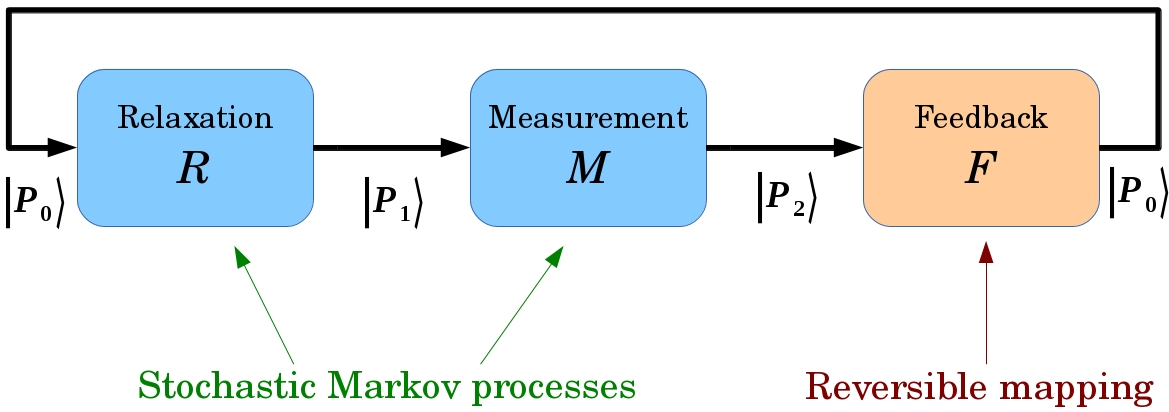}
\caption{Cycle of the engine. The boxes are represented by three $4\times 4$ transition matrices $\mathcal T_R$,$\mathcal T_M$, and $\mathcal T_F$. After many cycles, the probability distributions between the steps will become stationary.}
\label{fig:cycle}
\end{figure}

If the information engine runs repeatedly through many cycles, the probability distributions between the three steps will become stationary. The corresponding stationary probability distributions $|P_0\rangle$, $|P_1\rangle$, and $|P_2\rangle$ are represented as four-component vectors
\begin{equation}
|P_k\rangle=(P_k^{00},\,P_k^{01},\,P_k^{10},\,P_k^{11})^T, \quad (k=0,1,2) 
\end{equation}
are determined by the equations
\begin{eqnarray}
\label{SS}
\mathcal T_F\mathcal T_M\mathcal T_R \, |P_0\rangle &=& |P_0\rangle \nonumber \\
\mathcal T_R\mathcal T_F\mathcal T_M \, |P_1\rangle &=& |P_1\rangle \\
\mathcal T_M\mathcal T_R\mathcal T_F \, |P_2\rangle &=& |P_2\rangle \nonumber
\end{eqnarray}
with $|P_1\rangle=\mathcal T_R|P_0\rangle$ and $|P_2\rangle=\mathcal T_M|P_1\rangle$ (see Fig.~\ref{fig:cycle}).

The reduced stationary probability vectors of the system $(s)$ and the memory $(m)$ are given, respectively, as
\begin{eqnarray}
|P_k^{(s)}\rangle &=& (P_k^{00}+P_k^{01},\, P_k^{10}+P_k^{11})^T \nonumber \\
|P_k^{(m)}\rangle &=& (P_k^{00}+P_k^{10},\, P_k^{01}+P_k^{11} )^T\,.
\end{eqnarray}
Clearly, the measurement does not modify the system state, meaning that $|P_1^{(s)}\rangle =  |P_2^{(s)}\rangle$. Similarly, both the relaxation and feedback do not affect the memory, hence $|P_0^{(m)}\rangle =  |P_1^{(m)}\rangle$ and $|P_2^{(m)}\rangle =  |P_0^{(m)}\rangle$. As a result, in a stationary situation, the information acquired during the measurement is statistically the same in each cycle, implying that $|P_1^{(m)}\rangle =  |P_2^{(m)}\rangle$.

As a simple example, first consider the case of infinite-time relaxation and measurement ($t_R, t_M \rightarrow \infty$), where the matrices $\mathcal T_R$,$\mathcal T_M$, and $\mathcal T_F$ are given by Eqs.~(\ref{TRI}), (\ref{TMI}), and~(\ref{TFI}). In this case the normalized stationary probability vectors turn out to be given by
\begin{eqnarray}
\label{P0stat}
|P_0\rangle &=& \bigl(\bar \epsilon \bar q,\;\; \bar \epsilon q,\;\;\epsilon q,\;\;\epsilon \bar q\bigr)^T\\
\label{P1stat}
|P_1\rangle &=& \bigl( (\bar \epsilon \bar q + \epsilon q )\bar q ,\;\;
  (\bar \epsilon q  + \epsilon \bar q )\bar q ,\;\;
  (\bar\epsilon  \bar q + \epsilon q)q, \;\; (\bar \epsilon q+\epsilon  \bar q)q \bigr)^T \nonumber\\ \\
\label{P2stat}
|P_2\rangle &=& \bigl( \bar \epsilon \bar q ,\;\;\epsilon \bar q,\;\;\epsilon q,\;\; \bar \epsilon q\bigr)^T\,.
\end{eqnarray}
The corresponding reduced vectors for the system and the memory read
\begin{eqnarray}
\label{Pstat}
|P_0^{(s)}\rangle &=&  \begin{pmatrix}  \bar \epsilon \\ \epsilon \end{pmatrix} \nonumber  \\
|P_1^{(s)}\rangle =  |P_2^{(s)}\rangle &=& \begin{pmatrix} \bar q \\ q \end{pmatrix} \\
|P_0^{(m)}\rangle =  |P_1^{(m)}\rangle =|P_2^{(m)}\rangle &=&
  \begin{pmatrix} \bar \epsilon \bar q + \epsilon  q\\ \bar \epsilon q + \epsilon \bar q \end{pmatrix}. \nonumber
\end{eqnarray}
%
\section{Entropy and entropy production}
%
\begin{center} \textbf{1. Shannon entropy}\\ \end{center}
%
Given the stationary probability distributions $|P_k\rangle$, it is straightforward to compute the entropies of the system~$(s)$, the memory $(m)$, and the joint system $(sm)$ between the steps in a stationary cycle, using the definition of the Shannon entropy 
\begin{equation}
H = -\sum_c P^c \ln P^c \ ,
\end{equation}
where the sum runs over the vector components. Because of the aforementioned coincidence of various probability vectors we have $H_1^{(s)}=H_2^{(s)}$ and $H_1^{(m)}=H_2^{(m)}=H_3^{(m)}$. Furthermore, the reversibility of the feedback process guarantees that $H_2^{(sm)}=H_0^{(sm)}$.

For $t_R, t_M \rightarrow \infty$, the expressions for the Shannon entropies reduce to
\begin{eqnarray}
H_0^{(s)} &=& h(\epsilon)+h(\bar \epsilon) \nonumber\\
H_{1,2}^{(s)} &=& h(q)+h(\bar q) \\
H_{0,1,2}^{(m)} &=&  h(\epsilon\bar q+\bar \epsilon q)+h(\epsilon q + \bar \epsilon \bar q) \, ,\nonumber\\
H_{0,2}^{(sm)} &=& h(q)+h(\bar q)+h(\epsilon)+h(\bar \epsilon) \\
H_1^{(sm)} &=& h(q)+h(\bar q)+h(\epsilon\bar q+\bar \epsilon q)+h(\epsilon q + \bar \epsilon \bar q)\,,\nonumber
\end{eqnarray}
where we used the notation $h(p) := -p \ln p$.

During the relaxation process, where the system tries to restore the equilibrium distribution from the overpopulated ground state after energy extraction, the system entropy $H^{(s)}$ is expected to increase, provided that the error probability $\epsilon$ is sufficiently small ($\epsilon < q$). The same applies to the composite entropy $H^{(sm)}$. 

To summarize, the entropy changes during relaxation (R), measurement~(M), and feedback (F) are given by
\begin{eqnarray}
&& \Delta H_R^{(s)} > 0, \ \Delta H_M^{(s)} = 0, \ \Delta H_F^{(s)} = - \Delta H_R^{(s)} < 0 \ ,\nonumber\\
&& \Delta H_R^{(m)} = \Delta H_M^{(m)} = \Delta H_F^{(m)} = 0 \ ,\\
&& \Delta H_R^{(sm)} > 0, \ \Delta H_M^{(sm)} = -\Delta H_R^{(sm)}, \ \Delta H_F^{(sm)} = 0 \ .\nonumber
\end{eqnarray}
\vspace{3mm}

\begin{center} \textbf{2. Mutual information}\\ \end{center}
%
With these expressions, it is straightforward to compute the mutual information 
\begin{equation}
I_k = H^{(s)}_k + H^{(m)}_k - H^{(sm)}_k \geq 0,\quad (k=0,1,2) 
\end{equation}
which is a measure for the correlation between system and memory. This correlation is expected to build up during the measurement and then to decrease during feedback and relaxation, implying the inequalities
\begin{equation}
\Delta I_R < 0, \quad \Delta I_M > 0, \quad \Delta I_F < 0 \ .
\end{equation}
It is interesting to note that the change of the composite entropy is purely given by amount of mutual information acquired during the measurement, i.e.
\begin{equation}
\label{DHI}
\Delta H_M^{(sm)}=-\Delta I_M \quad \text{and} \quad \Delta H_R^{(sm)}=\Delta I_M \ .
\end{equation}
For $t_R, t_M \rightarrow \infty$, we have
\begin{eqnarray}
I_0 &=& h(\epsilon\bar q+\bar \epsilon q)+h(\epsilon q +\bar \epsilon \bar q)-h(q)-h(\bar q)\nonumber\\
I_1 &=& 0\\
I_2 &=& h(\epsilon\bar q+\bar \epsilon q)+h(\epsilon q+\bar \epsilon \bar q)-h(\epsilon)-h(\bar \epsilon) \ .\nonumber
\end{eqnarray}
Note that the result $I_1 = 0$ is true only if the relaxation time is infinite, which obviously destroys all correlations between system and memory. \\

\begin{center} \textbf{3. Entropy production}\\ \end{center}
%
Let us now turn to the entropy production. According to Schnakenberg~\cite{Schnakenberg,Gaspard,Seifert05}, whenever the system or the memory jumps spontaneously from the configuration $c$ to another configuration $c'$, the amount of entropy
\begin{equation}
\label{ep}
\Delta H^{env}_{c \to c'}(t)=\ln \frac{w_{c \to c'}(t)}{w_{c' \to c}(t)}
\end{equation}
is generated in the environment. Here, $w_{c \to c'}(t)$ denotes the transition rate at time $t$. Therefore, the mean entropy production rate is given by
\begin{equation}
\frac{\d}{\d t} H^{env} = \sum_{c \neq c'} P^c(t) w_{c \to c'}(t) \ln \frac{w_{c \to c'}(t)}{w_{c' \to c}(t)} \,.
\end{equation}
In an arbitrary nonequilibrium system, one would have to solve the master equation, plug the solution into the equation above, and integrate the resulting expression over a certain window of time. However, in the present case this is not necessary since the model is so simple that the rates happen to obey {\em detailed balance}, defined as $P^c_{stat}w_{c \to c'}=P_{stat}^{c'} w_{c'\to c}$ in the stationary equilibrium state. In this case, it is therefore straightforward to rewrite the equation given above as
\begin{eqnarray}
\frac{\d}{\d t} H^{env} &=&  \sum_{c \neq c'} \left( P^{c'}w_{c' \to c} -P^c w_{c \to c'}\right) \ln P^c_{stat} \nonumber \\
&=& \sum_{c} \frac{\d P^c}{\d t} \ln P^c_{stat} \,,
\end{eqnarray}
allowing us to compute the average entropy production in each step directly without integration by means of
\begin{equation}
\Delta H^{env} = \sum_c (P^c_{final}-P^c_{init}) \, \ln P^c_{stat}\,.
\label{H_env_general}
\end{equation}
Here $P^c_{init}$ and $P^c_{final}$ denote the initial and the final probabilities for a finite time span while $P^c_{stat}$ is the stationary probability distribution that would emerge after infinite long time. For example, during relaxation, $P^c_{stat}$ can be obtained by taking $t_R \to\infty$ in the expression for $P_1^c$ in Eq.~(\ref{P1stat}). Similarly, during measurement, $P^c_{stat}=\lim_{t_M\to\infty} P_2^c$ with $P_2^c$ given in Eq.~(\ref{P2stat}).

With the above formula, we obtain the following expressions for the entropy production in each process:
%
\begin{figure}
\centering\includegraphics[width=85mm]{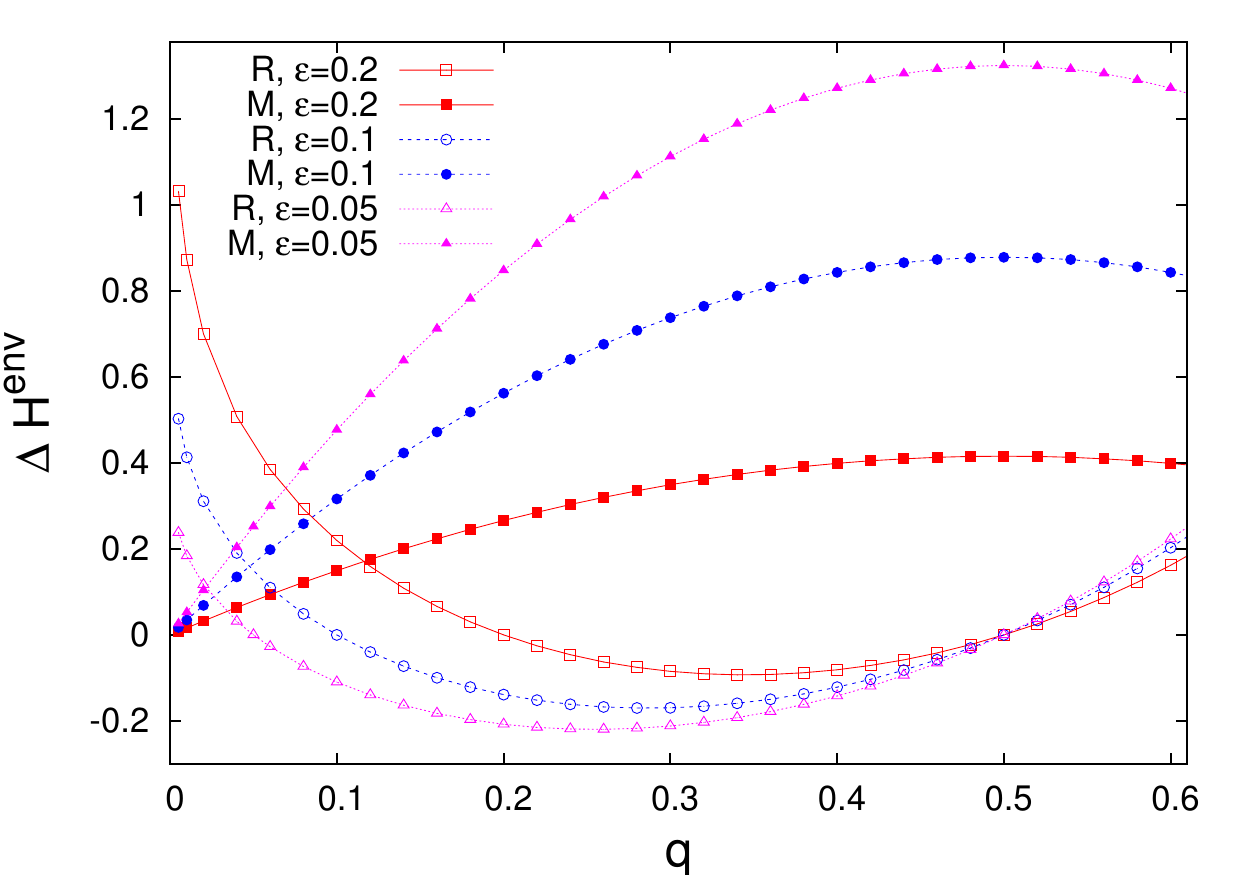}
\caption{Entropy production according to Eq.~(\ref{Henv_inf}) during relaxation (lower curves) and measurement (upper curves) for various values of $\epsilon$ as a function of $q$. If the measurement is accurate enough ($\epsilon<q <1/2$), the entropy production during relaxation becomes negative.}
\label{fig:entropyproduction}
\end{figure}
%
\begin{eqnarray}
\label{Henv}
\Delta H^{env}_R &=& ( P^{10}_0+P^{11}_0- P^{10}_1 -P^{11}_1 ) \ln \frac{ \bar q}{q} \nonumber\\
\Delta H_M^{env} &=& (P^{01}_1+P^{10}_1 -P^{01}_2-P^{10}_2) \ln \frac{\bar \epsilon}{\epsilon} \\
\Delta H_F^{env} &=& 0 \ .\nonumber
\end{eqnarray}
Note that these expressions hold for any finite $t_R$ and $t_M$. The last result is obvious since the feedback with $\mathcal T_F^2=\bf I$ is a reversible operation.

For $t_R, t_M \rightarrow \infty$, by inserting Eqs.~(\ref{P0stat})-(\ref{P2stat}) we get explicit expressions
\begin{eqnarray}
\label{Henv_inf}
\Delta H_R^{env} &=& -(q-\epsilon) \ln \frac{\bar q}{q} \\
\Delta H_M^{env} &=& 2q \bar q (\bar \epsilon- \epsilon) \ln\frac{\bar \epsilon}{\epsilon}\,, \nonumber
\end{eqnarray}
which are plotted for various error probabilities in Fig.~\ref{fig:entropyproduction}. As one can see, for $\epsilon<q<1/2$ the entropy production during relaxation $\Delta H_R^{env}$ is negative, meaning that the engine imports entropy (heat) from the environment rather than producing it. Obviously, this is the regime of interest where we would like to operate our information engine. However, as can be seen, the negative entropy production during relaxation is always overcompensated by a positive one during the measurement, which is consistent with the second law of thermodynamics. Notice that the entropy production during measurement is always positive since $\epsilon <1/2$.

\section{Work extraction and supply}
\begin{figure}
\centering\includegraphics[width=85mm]{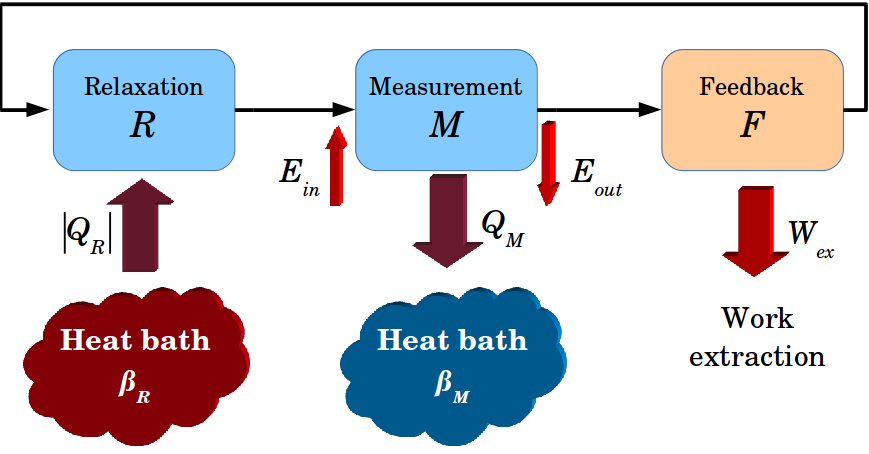}
\caption{Balancing flow of heat and work during the engine cycle (see text). }
\label{fig:totalscheme}
\end{figure}

%
By virtue of Clausius' law $\d Q = T \d H$, the produced entropy can be translated directly into an amount of heat. In most studies, it is usually assumed that the temperatures of the system and the memory are identical.
However, for the sake of generality, let us allow the temperatures to be different, assigning $\beta_R=1/T_R$ during relaxation and $\beta_M=1/T_M$ during measurement, as sketched in Fig.~\ref{fig:totalscheme}. Thus the respective heat contributions averaged over all possible stochastic trajectories are given by
\begin{equation}
\langle  Q_R \rangle
=\beta_R^{-1} \Delta H^{env}_R\,,\qquad
\langle  Q_M \rangle =\beta_M^{-1} \Delta H^{env}_M\,.
\label{heat_entropy}
\end{equation}
Here we use the sign convention that heat flowing away from the engine into the environment has a positive sign, i.e. we expect  $\langle  Q_R \rangle$ to be negative and $\langle  Q_M \rangle$ to be positive.

In order to maintain stationarity of the system after one engine cycle, the average work $\langle W_{ex} \rangle $ extracted during feedback should exactly balance the average heat $\langle Q_R\rangle$ during relaxation, i.e.
\begin{equation}
\label{WW1}
\langle W_{ex} \rangle = -\langle Q_R \rangle >0 \,.
\end{equation}
Consequently, the measurement process does not change the energy of the system. This requires an additional influx of energy in the form of extra work $\langle W_{sup} \rangle$ into the memory which is necessary to compensate the loss of heat $\langle Q_M\rangle $ flowing away to the environment during the measurement process:
\begin{equation}
\label{WW2}
\langle W_{sup} \rangle = \langle Q_M \rangle >0 \, .
\end{equation}
The average \textit{net} work performed by the machine, defined as the difference of extracted and supplied work, is therefore given by
\begin{equation}
\langle W_{net} \rangle = \langle W_{ex}\rangle -\langle W_{sup}\rangle =-\langle Q_R\rangle -\langle Q_M \rangle\ .
\end{equation}
Note that the net work can change its sign depending on the choice of the parameters $q$, $\epsilon$, $t_R$, and $t_M$.

Let us first compute the extracted work. According to Sect. II.3 $\langle W_{ex} \rangle $ is given by $(P_2^{11}-P_2^{01})\Delta E$, where $\Delta E=\beta_R^{-1} \ln \frac{ \bar q}{q}$, see~(\ref{DeltaE}). Using $|P_1^{(s)}\rangle =  |P_2^{(s)}\rangle$ (no system change during measurement) together with the feedback identities $P_2^{01}=P_0^{11}$ and $P_2^{10}=P_0^{10}$ (flip $s$ only when $m=1$), it is easy to show explicitly that
\begin{equation}
\label{Wex1}
\langle W_{ex} \rangle = - \beta_R^{-1} \Delta H_R^{env} = -\langle Q_R\rangle.
\end{equation}
The additional work $\langle W_{sup} \rangle$ supplied during the measurement process can be interpreted as the energy needed to operate the measurement device. Technically this contribution comes from the fact that the energy levels of the joint system are different during relaxation and measurement so that extra energy is needed to move them around. For example, this could be done by applying an external potential in order to make the energy level of the erroneous composite state $(s,1-s)$ higher than that of the correctly measured state $(s,s)$ by the amount of $\Delta E^\prime=\beta_M^{-1} \ln \frac{ \bar \epsilon}{\epsilon}$. When the external potential is turned on just before the measurement, the average energy of the composite of system and memory increases by $\langle E_{in} \rangle$ and similarly it looses the energy $\langle E_{out} \rangle$ when the potential is turned off at the end of the measurement:
\begin{equation}
\langle E_{in} \rangle = \frac{P^{01}_1 + P^{10}_1}{\beta_M}\ln \frac{\bar{\epsilon} }{\epsilon} \ , \quad
\langle E_{out} \rangle =
\frac{P^{01}_2 + P^{10}_2}{\beta_M}\ln \frac{\bar{\epsilon} }{\epsilon} \ .
\end{equation}
Comparing the difference $\langle W_{sup} \rangle = \langle E_{in} \rangle - \langle E_{out} \rangle$ with Eq.~(\ref{Henv}) one can see immediately that 
\begin{equation}
\label{Wex2}
\langle W_{sup} \rangle =  \beta_M^{-1} \Delta H_M^{env} = \langle Q_M\rangle\,. 
\end{equation}
In the limit $t_R, t_M \rightarrow \infty$, the average work contributions read
\begin{eqnarray}
\label{WW_inf}
\langle W_{ex} \rangle &=& \frac{q-\epsilon}{\beta_R} \ln \frac{\bar q}{q} \\
\langle W_{sup} \rangle &=& \frac {2q \bar q (\bar \epsilon- \epsilon)}{\beta_M} \ln\frac{\bar \epsilon}{\epsilon}. \nonumber
\end{eqnarray}
%

\section{Thermodynamic second laws and fluctuation theorems}
\begin{figure}
\centering\includegraphics[width=85mm]{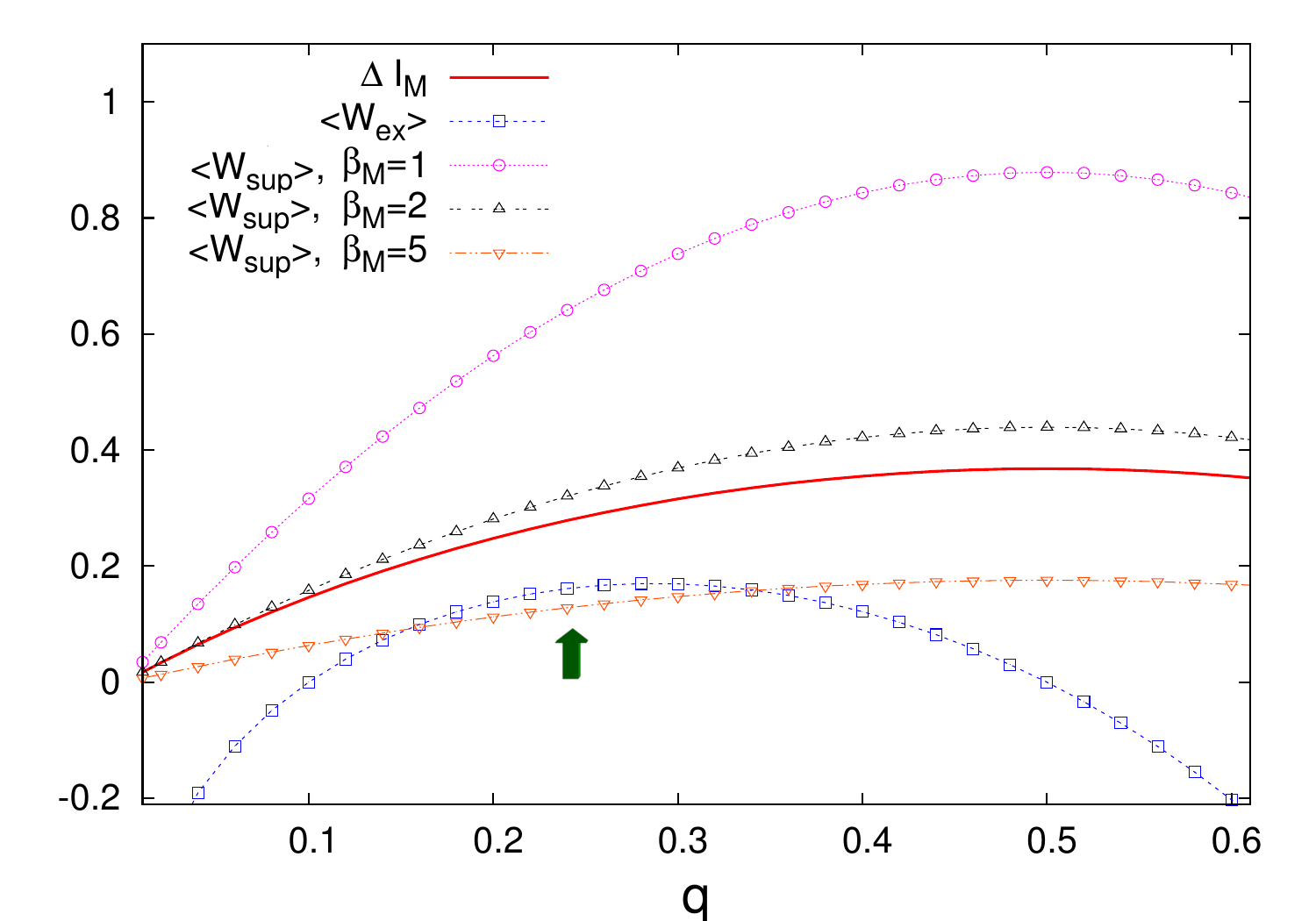}
\caption{Infinite-time relaxation and measurement. The extracted work $\langle W_{ex} \rangle$, supplied work $\langle  W_{sup} \rangle$, and the mutual information $\Delta I_M= I_2$ are displayed as functions of $q$ for a constant error probability  $\epsilon =0.1$. Here we choose $\beta_M =1, 2,$ and 5 with $\beta_R =1$. Note that $\beta_M \langle  W_{sup} \rangle = \Delta H^{env}_M $ is independent of $\beta_M$. $\langle W_{ex} \rangle$ is positive at $\epsilon < q< 1/2$ as expected from Eq.~(\ref{WW_inf}). When the temperature of the measurement reservoir is sufficiently lower than that of the relaxation reservoir, for example $\beta_M=5$, the extracted work $\langle W_{ex} \rangle$ can be larger than the supplied work $\langle  W_{sup} \rangle$, as indicated by the arrow.
}
\label{fig:IFT_infinite}
\end{figure}

%
The total entropy production of the whole setup during relaxation (R) and measurement (M) is given by
\begin{eqnarray}
\Delta H^{tot}_R &=& \Delta H_R^{(sm)} + \Delta H^{env}_R \\
\Delta H^{tot}_M &=& \Delta H_M^{(sm)} + \Delta H^{env}_M \, .
\end{eqnarray}
According to Eq.~(\ref{DHI}), the entropy differences in the joint system are given solely in terms of the mutual information difference
\begin{equation}
\Delta H^{sm}_{R} = -\Delta H^{sm}_{M} = \Delta I_M
\end{equation}
while the entropy differences in the environment are given in terms of the transferred heat by $\Delta H^{env}_{R,M}=\beta_{R,M}\langle Q_{R,M}\rangle$. Using Eqs.~(\ref{Wex1}) and (\ref{Wex2}) we arrive at
\begin{eqnarray}
\Delta H^{tot}_R &=& 
+\Delta I_M - \beta_R \langle  W_{ex}\rangle  \;\;\; \\
\Delta H^{tot}_M &=& 
-\Delta I_M + \beta_M \langle  W_{sup} \rangle \nonumber\,.
\end{eqnarray}
For the total system including the environment the second law of thermodynamics should be satisfied for each process, i.e.,
\begin{equation}
\Delta H^{tot}_R \ge 0 \quad \text{and} \quad \Delta H^{tot}_M \ge 0 \ ,
\label{Thermod_lawsH}
\end{equation}
or equivalently
\begin{equation}
\beta_R \langle  W_{ex} \rangle \ \leq \Delta I_M  \leq \ \beta_M \langle  W_{sup}\rangle \ .
\label{Thermod_laws}
\end{equation}
Most existing studies are only interested in the first inequality during relaxation, while the other one during measurement is ignored. The purpose of this work is to point out that there is a \textit{second} inequality for the measurement process as well, and that the two inequalities are complementary with respect to each other. More specifically, if $\beta_R=\beta_M$, the extracted work is bounded from above by the mutual information while the work required to operate the memory is bounded from below by the same threshold. This means that the setup cannot be used to gain work out of nothing, $\langle W_{net} \rangle \le 0$, as expected by the second law. However, if the two reservoir temperatures were different $(\beta_R<\beta_M)$, the extracted work could be larger than the supplied one, in which case the system operates like a conventional heat engine (see Fig.~\ref{fig:IFT_infinite}).

It is almost trivial to construct the integral fluctuation theorems (IFT's) through the standard approach of stochastic thermodynamics~\cite{Seifert05} by considering the heat along all possible trajectories in the composite configurational state space. With an appropriate definition of the Shannon entropy for a given trajectory~\cite{Seifert05}, one can easily get the fluctuation theorems for the total entropy production for each process as
\begin{equation}
\langle e^{-\Delta H^{tot}_{R}(traj.)}\rangle=1 \,, \quad \langle e^{-\Delta H^{tot}_{M}(traj.)}
\rangle=1 \, .
\end{equation}
The second laws in Eq.~(\ref{Thermod_lawsH}) are simple consequences of the IFT's with $\Delta H^{tot}_{R,M}=\langle {\Delta H^{tot}_{R,M}(traj.)}\rangle$. It is rather tricky to find the IFT's in terms of work and mutual information, because it requires an equilibrium state as an initial condition. This is the case only when $t_R$ becomes infinite so that the system is in equilibrium at the start of the measurement as well as at the beginning of the feedback. However, note that the bounds for works in Eq.~(\ref{Thermod_laws}) are valid even if $t_R$ is finite.

\section{Finite-Time relaxation and measurement}

In practice, an engine is only useful if the cycle time $\tau$ is finite. Thus, it is obviously of interest to derive all physical quantities as function of the cycle time. This allows one to find the optimum for maximal power generation, as will be discussed in the next section.

It is straightforward to obtain the transition matrices for relaxation and measurement for finite time spans $t_R$ and $t_M$:
\small
\begin{equation}
\label{T_finite}
\mathcal T_R(t_R) \;=\; \begin{pmatrix}
\mathcal R+ \bar{\mathcal R}\bar q & 0 & \bar{\mathcal R} \bar q & 0 \\
0 & \mathcal R+ \bar{\mathcal R}\bar q & 0 &  \bar{\mathcal R} \bar q \\
 \bar{\mathcal R}q & 0 &  \mathcal R +\bar{\mathcal R} q & 0 \\
0 &\bar{\mathcal R} q & 0 & \mathcal R+\bar{\mathcal R}q
\end{pmatrix}
\end{equation}
\begin{equation}
\mathcal T_M(t_M) \;=\; \begin{pmatrix}
\mathcal M+ \bar{\mathcal M} \bar \epsilon &    \bar{\mathcal M} \bar \epsilon& 0 & 0 \\
 \bar{\mathcal M} \epsilon & \mathcal M+  \bar{\mathcal M} \epsilon &0 &0 \\
0&0& \mathcal M+ \bar{\mathcal M}\epsilon &   \bar{\mathcal M}\epsilon \\
0 &0 &  \bar{\mathcal M} \bar \epsilon & \mathcal M+  \bar{\mathcal M} \bar \epsilon
\end{pmatrix} \nonumber ,
\end{equation}
\normalsize
where
\begin{eqnarray}
\mathcal R := e^{-t_R}\  \quad &\text{and}& \quad \bar{ \mathcal R} := 1-\mathcal R \\
\mathcal M := e^{-t_M}\ \quad &\text{and}& \quad \bar{ \mathcal M} := 1-\mathcal M \,.
\end{eqnarray}
Note that the transition probability from $(s,1-s)$ to $(s,s)$ during measurement, $\bar{\mathcal M} \bar \epsilon$, is smaller than $\bar \epsilon$, which means that the measurement for finite $t_M$ is less accurate than in the limit of infinite time. 

\begin{figure}
\centering\includegraphics[width=85mm]{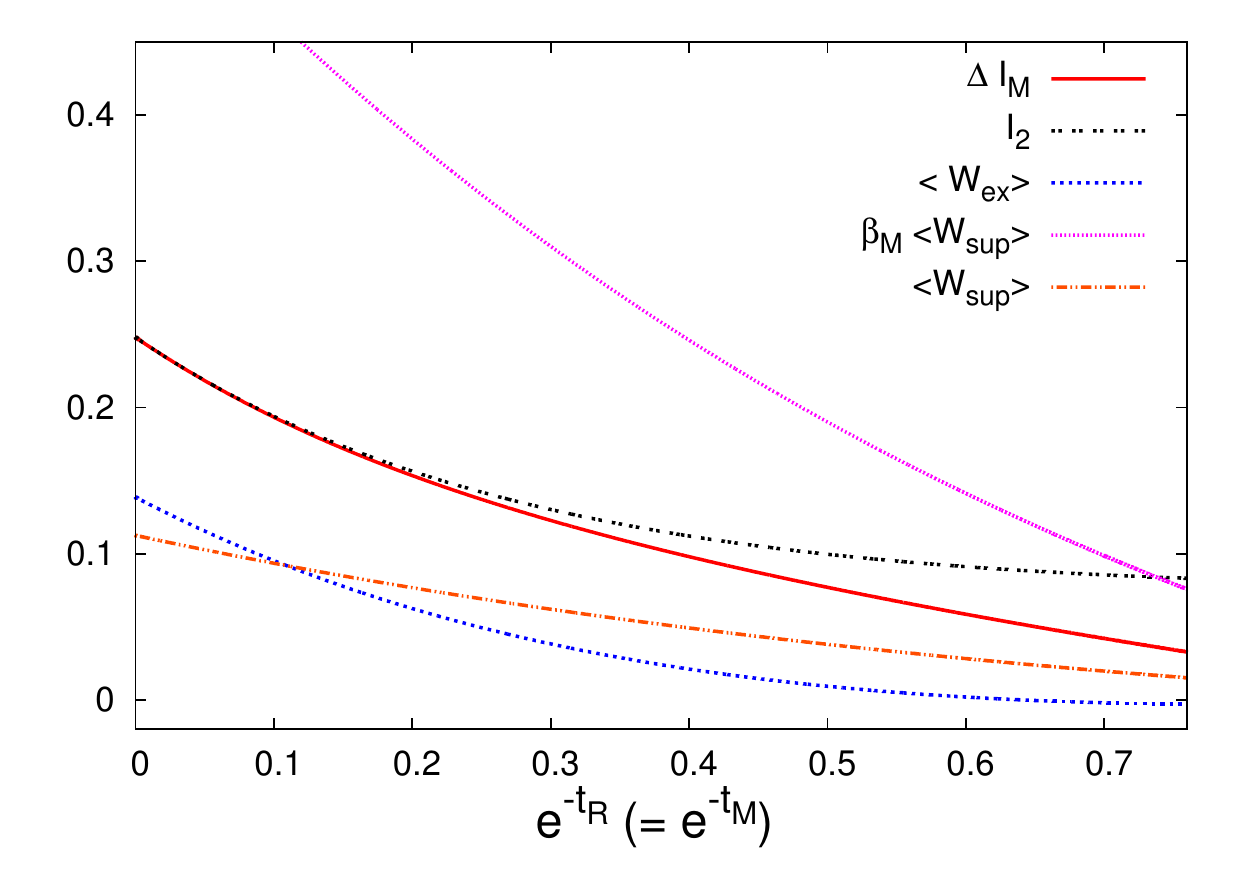}
\caption{Finite-time relaxation and measurement. The extracted work $\langle W_{ex} \rangle $, supplied work $\langle W_{sup} \rangle$, and the mutual information $I_2$ and $\Delta I_M$ are shown as functions of
$e^{-t_R}$. Here, we choose $t_R = t_M$, $q=0.2$, $\epsilon =0.1$, $\beta_R =1$, and $\beta_M =5$.
}
\label{fig:IFT_finite}
\end{figure}

Solving Eq.~(\ref{SS}), one can find explicit but complicated expressions for all stationary distributions such as $|P_0\rangle$, $|P_1\rangle$, and $|P_2\rangle$ for finite $t_R$ and $t_M$, which are not shown here explicitly. The heat dissipation during the finite-time relaxation and measurement can be obtained from Eq.~(\ref{Henv}) while the extracted work and the supplied work are given by Eqs.~(\ref{WW1}) and (\ref{WW2}).

Using the relation $|P_2 \rangle = \mathcal T_M |P_1 \rangle$ we find that
\begin{equation}
\langle W_{sup} \rangle
= \frac{ \mathcal E -\epsilon}{\beta_M} \bar{\mathcal M} \ln \frac{\bar \epsilon}{\epsilon}\ .
\label{eq:finite_Ws}
\end{equation}
Here $\mathcal E \equiv P_1^{01}+P_1^{10}$ is explicitly given by
\begin{equation}
\mathcal E (t_R, t_M) = \frac{ \bar {\mathcal R} ( \bar \epsilon q +\epsilon \bar q)  + \alpha \left ( \bar{\mathcal R} q +  \mathcal R  \bar{\mathcal M} \epsilon \right )}
{1-\alpha \mathcal R \mathcal M},
\label{eq:cross}
\end{equation}
with $\alpha = \mathcal R + \bar{\mathcal R}(\bar q- q) (\bar \epsilon - \epsilon)$. In a similar manner, we obtain $\langle W_{ex} \rangle$ as the function of $\mathcal E$:
\begin{equation}
\langle W_{ex} \rangle= \frac{q-\epsilon -\mathcal M (\mathcal E - \epsilon) }{\beta_R} \bar{\mathcal R }\ln \frac{ \bar q}{q}\,.
\label{eq:finite_Wex}
\end{equation}
In the limit of $t_R, t_M\to\infty$ ($\mathcal R, \mathcal M\to 0$), we consistently recover Eq.~(\ref{WW_inf}).

Note that the finite-time works in~(\ref{eq:finite_Ws}) and (\ref{eq:finite_Wex}) decrease monotonously with $\mathcal R$ and $\mathcal M$ (see Fig.~\ref{fig:IFT_finite}). Moreover, since the correlation between system and memory builds up continuously during the measurement process, it is obvious that $\Delta I_M$ decreases with $\mathcal M$, remaining positive by definition. The positivity of $\Delta I_M$ guarantees that $\langle W_{sup}\rangle$ is also positive. On the other hand, $\langle W_{ex}\rangle$ can be negative for short-time measurement and relaxation, as its upper bound $\Delta I_M$ approaches zero for $t_M\to 0$. 

The monotonous dependence shown in Fig.~\ref{fig:IFT_finite} suggests that $\langle W_{ex}\rangle$ becomes maximal for maximal measurement accuracy ($t_M\to\infty$) and full relaxation ($t_R\to\infty$) in order to redistribute and pump the overpopulated ground state $s=0$ back to the energetically excited state $s=1$. Therefore, both limit $t_M,t_R\to\infty$ have to be carried out simultaneously. To establish this combined limit conveniently, let us from now on set 
\begin{equation}
t_R=t_M=\tau/2\,,
\end{equation}
meaning that $\mathcal R= \mathcal M = e^{-\tau/2}$. With this convention we expect $\langle W_{ex}\rangle$ to be maximal in the limit of infinite cycle time ($\tau \to \infty$). Moreover, as $\tau$ decreases, we expect  $\langle W_{ex}\rangle$ to decrease and eventually to become negative.

If $\langle W_{ex} \rangle > \langle W_{sup} \rangle$, the system operates like a conventional heat engine. For infinite $\tau$ the net work $\langle W_{net} \rangle$ is maximal, but the power (net work per unit time) vanishes. For finite but sufficiently large $\tau$ and properly chosen parameters the system still produces a positive net work, hence the power is positive. However, as can be seen in Fig.~\ref{fig:IFT_finite}, the curves for $\langle W_{ex} \rangle$ and $\langle W_{sup} \rangle$ cross each other at some finite cycle time $\tau=\tau_s$. At this point we do no longer obtain any net work from the engine, hence the power vanishes again. Consequently, there will be a particular cycle time in between, at which the power of the engine is maximal. In the next section, we will discuss this aspect in more detail.

\section{Efficiency}

\begin{figure}
\centering\includegraphics[width=55mm]{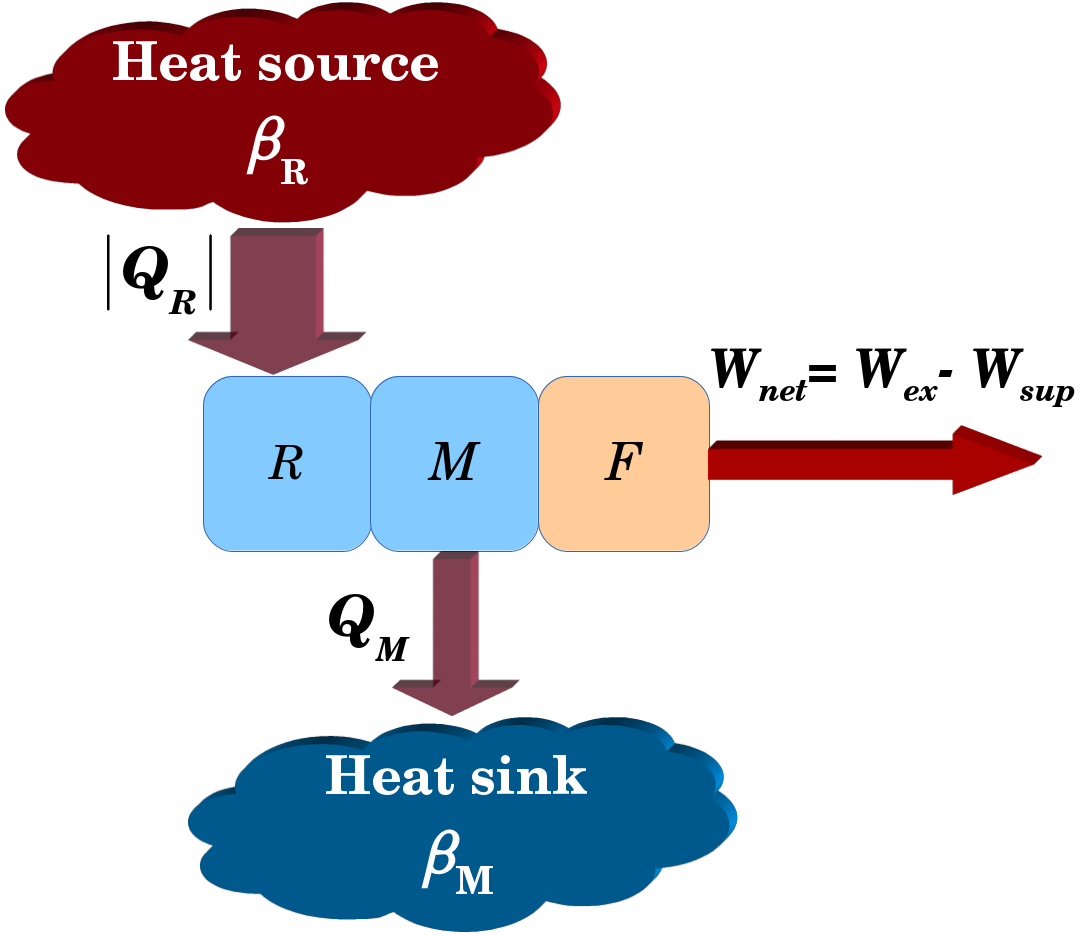}
\caption{Interpretation of the information engine as a conventional heat engine. Heat $|\langle Q_R \rangle|$  flows from
the heat reservoir at $\beta_R$ (high temperature) into the engine which produces the work gain $\langle W_{net} \rangle =
\langle W_{ex} \rangle - \langle W_{sup} \rangle$. The remaining heat $\langle Q_M \rangle$ is transferred to another reservoir
at $\beta_M$ (low temperature).
 }
\label{fig:totalscheme2}
\end{figure}
\begin{figure}
\centering\includegraphics[width=85mm]{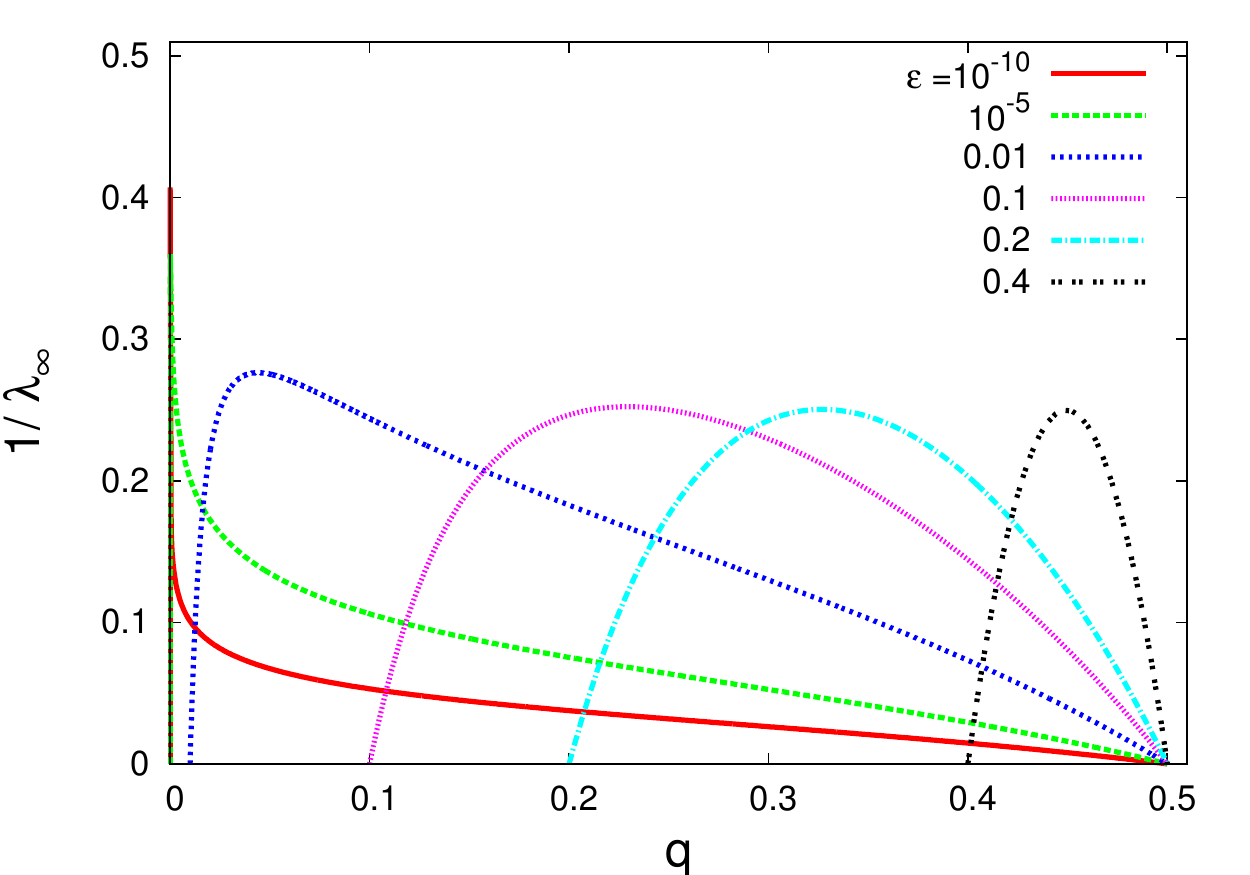}
\caption{$1/\lambda_{\infty}$ according to Eq.~(\ref{eq:lambda_inf}) as function of $q$ for several values of $\epsilon$.
In the infinity-time limit, $\langle W_{ex} \rangle$  and $1/\lambda_{\infty}$
are positive only for $ \epsilon \leq q \leq 1/2$. Note that, as $\epsilon$ goes to
zero, $\lambda_{\infty}$ approaches to 2, the minimum, at $q \to 0$.
}
\label{fig:eff_infinite}
\end{figure}

%
Let us now assume that the information engine operates in a regime where the net work is positive. In this case the whole setup can be interpreted as a conventional heat engine, as sketched in Fig.~\ref{fig:totalscheme2}. As $\beta_R<\beta_M$, the upper reservoir for the relaxation process plays a role of a high-temperature heat source, while the lower reservoir in contact with the memory device acts as a heat sink. The \textit{efficiency} of this heat engine in a single cycle is defined in the usual way as
\begin{equation}
\eta(\tau) = \frac{\langle W_{net} \rangle}{|\langle Q_{R} \rangle|}
     =1- \frac{\langle W_{sup} \rangle}{\langle W_{ex} \rangle }.
\label{eq:eff1}
\end{equation}
Using Eqs.~(\ref{eq:finite_Ws}) and (\ref{eq:finite_Wex}) the efficiency can be rewritten as
\begin{equation}
\eta(\tau)
=1 - \frac{\beta_R}{\beta_M}  \lambda (\tau),
\label{eq:eff2}
\end{equation}
where
\begin{equation}
\lambda (\tau)=  \frac{( \mathcal E - \epsilon)  \bar{\mathcal M }\ln \left( \bar \epsilon/ \epsilon \right) }
{ \left[ q-\epsilon -\mathcal M (\mathcal E - \epsilon) \right ] \bar{\mathcal R }\ln \left( \bar q /q \right )}.
\end{equation}
Note that the relaxation and measurement processes are not quasi-static so that even in the limit $\tau\to\infty$ the engine never reaches the Carnot efficiency $\eta_c=1-\frac{\beta_R}{\beta_M}$. Instead, we find that the efficiency is limited by a different upper bound $\eta_{max}$ which can be computed as follows. According to the monotonicity arguments discussed in the preceding section, $\eta(\tau)$ is expected to become maximal in the limit $\tau \to \infty$. This suggests that
\begin{equation}
\eta(\tau) \;\leq\; \eta_{max} \equiv \lim_{\tau \to \infty} \eta(\tau)= 1-\frac{\beta_R}{\beta_M} \lambda_{\infty},
\end{equation}
where
\begin{equation}
\lambda_{\infty}= \lim_{\tau \to \infty} \lambda(\tau) = \frac{ 2q\bar q (\bar \epsilon -\epsilon) \ln \left( \bar \epsilon/\epsilon \right ) }
{(q-\epsilon) \ln \left( \bar q/q \right) }\,.
\label{eq:lambda_inf}
\end{equation}
Fig.~\ref{fig:eff_infinite} shows $1/\lambda_{\infty}$ as function of $q$ for several values of~$\epsilon$. It is obvious that $1/\lambda_{\infty}$ is positive for $\epsilon \leq q \leq 1/2$ with a unimodal shape due to the similar behavior of $\langle W_{ex} \rangle$ demonstrated in Fig.~\ref{fig:IFT_infinite}. As $\epsilon\to 0$, $1/\lambda_{\infty}$ approaches zero except for $q\approx 0$. In the limit of both $q \to 0$ and $\epsilon \to 0$, $\lambda_{\infty}$ approaches a constant bounded from below by $\lambda_{\infty}\ge 2$. Consequently, in order to obtain a positive work gain, the difference of temperatures should be at least $\beta_M /\beta_R \geq 2$ for the infinite-time process. In short, we find that the efficiency of the information engine is bounded by
\begin{equation}
\eta_{max} \,\leq\, 1- \frac{2\beta_R}{\beta_M}  \,< \, \eta_c \ .
\nonumber
\end{equation}
%
%
In conventional heat engines operating with a finite cycle time, thermodynamic processes are no longer quasi-static, leading to an efficiency below the Carnot bound $\eta_c$. In our case, we also find that $\eta(\tau)$ becomes maximal in the limit $\tau\to\infty$. However, in contrast to the Carnot limit, where the entropy production vanishes, the entropy production per cycle in our model $\Delta H^{tot}_R + \Delta H^{tot}_M$ becomes \textit{also maximal} at $\tau =\infty$. This is one of the crucial features of our information engine which is totally distinct from conventional heat engines. Nevertheless, the entropy production \textit{rate} (per unit time) decreases with increasing $\tau$ and finally vanishes at $\tau=\infty$. Hence, one may also say that the maximum efficiency is found at the minimum entropy production rate.

\begin{figure}
\centering\includegraphics[width=85mm]{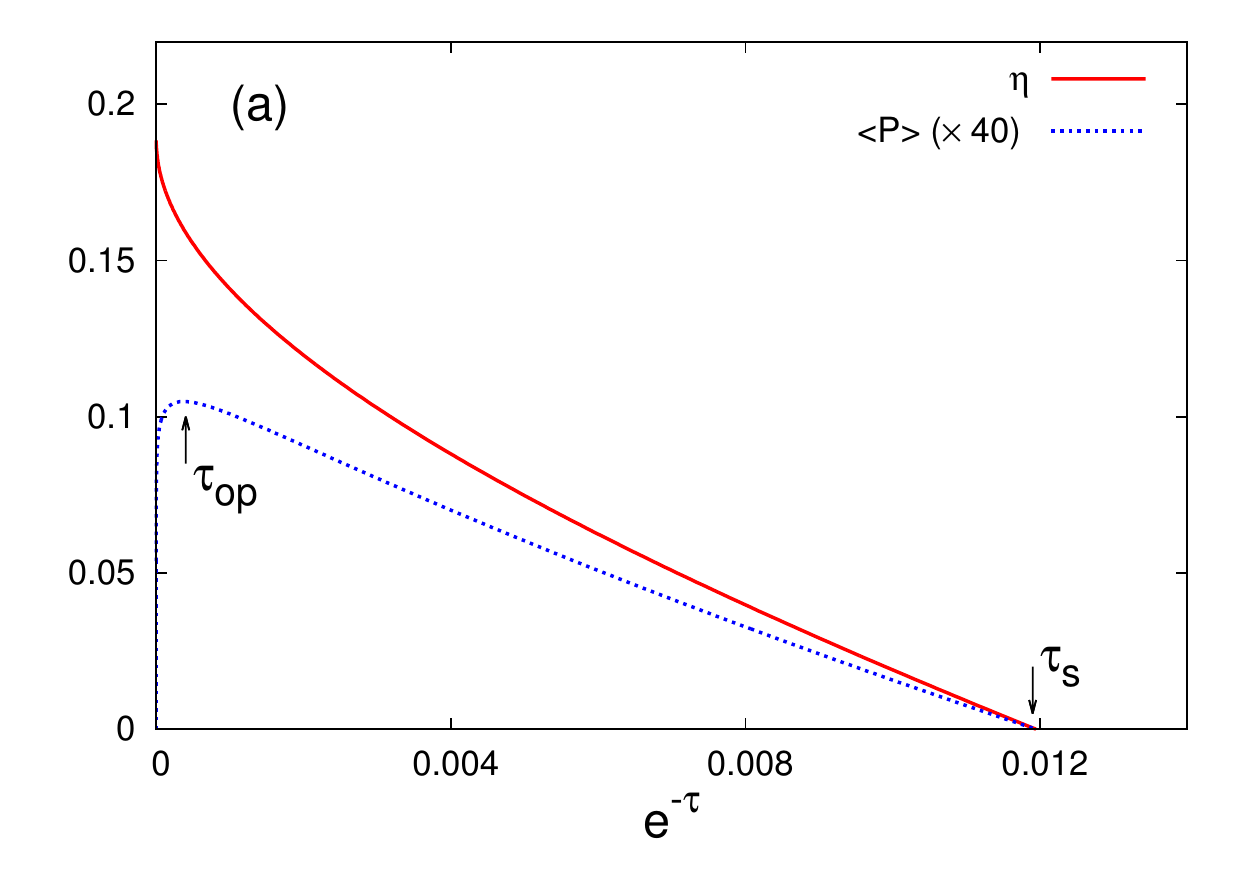}
\centering\includegraphics[width=88mm]{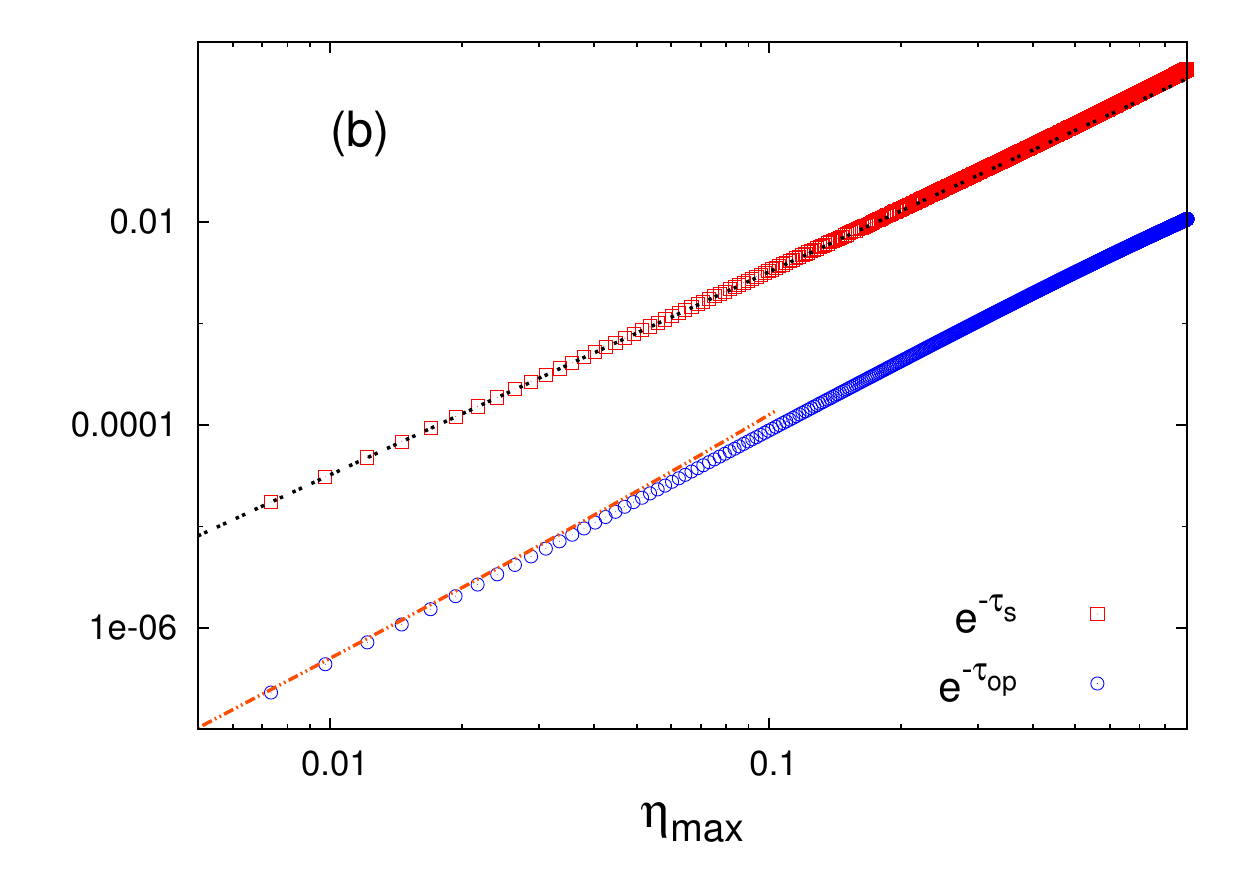}
\vspace*{-5mm}
\caption{ 
(a) The efficiency $\eta$ and power gain $\langle P \rangle$ as functions of $e^{-\tau }$.
(b) $e^{-\tau_s}$ and $e^{-\tau_{op}}$  as functions of $\eta_{max}$. In (b), the dotted line $\sim x^2$ is obtained from Eq.~(\ref{eq:tau_s}) and the dashed line corresponds to Eq.~(\ref{eq:tau_op}). We choose $t_R = t_M$, $q=0.2$, $\epsilon =0.1$, and $\beta_R =1$ for both (a) and (b). We set $\beta_M =5$ for (a) and vary $\beta_M$ for (b).
}
\label{fig:eff_finite}
\end{figure}

Similarly, the average power gain, $\langle P \rangle \equiv \langle W_{net} \rangle/\tau$ in fact vanishes for $\tau \to \infty$ because $\langle W_{net} \rangle$ remains finite in this limit. For a realistic engine, we usually want to optimize the power gain, trading off the efficiency against the cycle time. As expected, $\langle P \rangle $ is maximized at some {\em finite} time, $\tau_{op}$ between $\tau_s$ and $\infty$ as shown in Fig.~\ref{fig:eff_finite}~(a).

\begin{figure}
\centering\includegraphics[width=85mm]{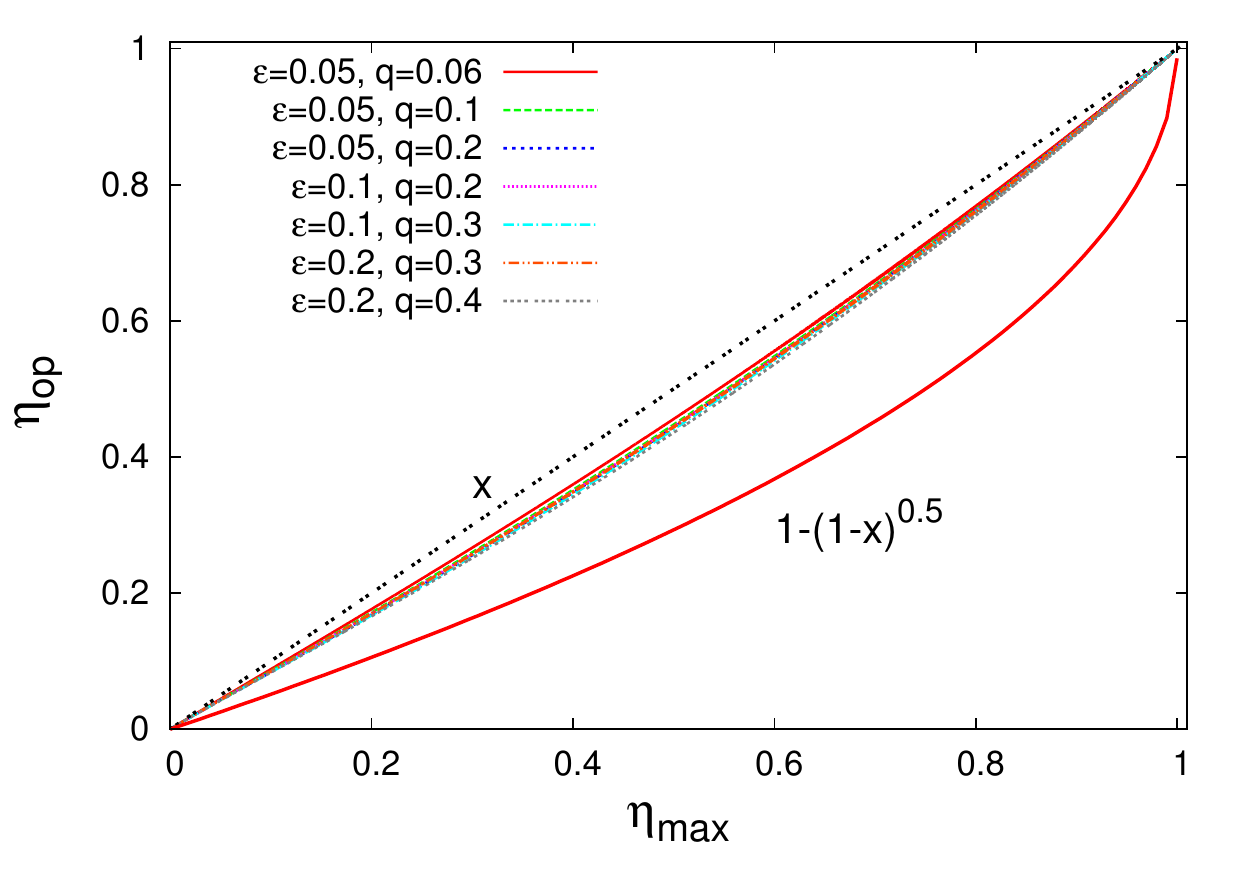}
\vspace*{-3mm}
\caption{The efficiency $\eta_{op}$ at the optimal power for various $\epsilon$ and $q$. Here we choose $k=k'$ and $t_R = t_M$ for simplicity.}
\label{fig:eff_maxp}
\end{figure}

The efficiency of heat engines at maximal power has been studied previously in Refs.~\cite{eff_CA,eff_Broeck,eff_Esposito}. Especially, for the Curzon-Ahlborn (CA) endoreversible model~\cite{eff_CA}, it is well-known that the efficiency $\eta_{\rm CA}$ at the optimal power is given by $\eta_{\rm CA}= 1- \sqrt{1-\eta_c }$. In Fig.~\ref{fig:eff_maxp}, we plot the efficiency at optimal power $\eta_{op}=\eta(\tau=\tau_{op})$ according to Eq.~(\ref{eq:eff2}) as a function of $\eta_{max}$ instead of $\eta_c$. It turns out that the functional behavior of $\eta_{op}$ is completely different from $\eta_{\rm CA}$.

%
%

In more general situations, it has been found that the efficiency at the maximum power obeys a universal form, $\eta_{op} = \eta_c /2 + O(\eta_c^2)$ for small $\eta_c$, when the engine and heat baths are strongly coupled~\cite{eff_Broeck,eff_Esposito}. Our information engine exhibits a completely distinct behavior even for small $\eta_{max}$. As seen in Fig.~\ref{fig:eff_maxp}, $\eta_{op}$ is more or less the same as $ \eta_{max}$ in this regime.

In order to investigate this unusual behavior in more detail, we now  examine $\eta_{op}$ analytically for small $\eta_{max}$. As the stall time $\tau_s$ and thus $\tau_{op}$ ($>\tau_s$) becomes large, a small $\mathcal R$ expansion may be valid for small $\eta_{max}$. Expanding $\eta(\tau)$ in Eq.~(\ref{eq:eff2}) up to the linear order of $\mathcal R=e^{-\tau/2}$, we get
\begin{equation}
\eta (\tau) \approx
\eta_{max} -\left( \frac{\mathcal E_1}{\mathcal E_0 -\epsilon }
+ \frac{\mathcal E_0 - \epsilon}{q-\epsilon} \right ) e^{-\tau/2},
\label{eq:expand1}
\end{equation}
where we have used $\mathcal E \approx \mathcal E_0 + \mathcal E_1 \mathcal R$ with
\begin{eqnarray}
\label{eq:E01}
\mathcal E_0 &=& 2 q \bar q (\bar \epsilon - \epsilon) + \epsilon \\
\mathcal E_1 &=& (q-\epsilon) - (\bar q- q)(\bar \epsilon -\epsilon) (q -\epsilon) - 2 q \bar q (\bar \epsilon - \epsilon).
\nonumber
\end{eqnarray}
We also checked that the linear coefficient inside the parentheses in Eq.~(\ref{eq:expand1}) is always numerically positive in the interval $\epsilon <q<1/2$.

As the stall time $\tau_s$ is defined by $\eta(\tau_s)=0$, Eq.~(\ref{eq:expand1}) immediately gives us
\begin{equation}
e^{-\tau_s} \approx  A_s \, \eta^2_{max},
\label{eq:tau_s}
\end{equation}
where the coefficient $A_{s}$ is given by
\begin{equation}
A_s = \left( \frac{\mathcal E_1}{\mathcal E_0 -\epsilon } + \frac{\mathcal E_0 - \epsilon}{q-\epsilon} \right ) ^{-2}\,.
\end{equation}
It turns out that the scaling behavior in Eq.~(\ref{eq:tau_s}) extends quite well to finite $\eta_{max}$, as can be seen in Fig.~\ref{fig:eff_finite} (b).

The optimal time $\tau_{op}$ should satisfy
\begin{equation}
\left. \frac{d \langle P \rangle } {d \tau } \right |_{\tau_{op}} = 0\, , \nonumber
\end{equation}
which is rewritten as
\begin{equation}
\frac{\langle  W_{net} (\mathcal R_{op}) \rangle }{\tau_{op}} =
\left. \frac{d \langle W_{net} (\mathcal R) \rangle }{d \tau } \right |_{\tau_{op}},
\label{eq:dpdt}
\end{equation}
where $\mathcal R_{op} = e^{-\tau_{op}/2}$. As $\tau_{op}>\tau_s$, $\mathcal R_{op}$ should be also small. This allows us to expand the above equation for small $\mathcal R_{op}$, yielding
\begin{equation}
 1- \frac{\beta_R}{\beta_M} \lambda_{op} \;=\;
\frac{\tau_{op}}{2} \left( \frac{\mathcal E_1}{\mathcal E_0 -\epsilon }
+ \frac{\mathcal E_0 - \epsilon}{q-\epsilon} \right ) \mathcal R_{op},
\label{eq:dpdt_R}
\end{equation}
where $\mathcal E_0$ and $\mathcal E_1$ are the same as in Eq.~(\ref{eq:E01}), and $\lambda_{op}$ is given by
\begin{equation}
\lambda_{op} = \lambda_{\infty} + \lambda_{\infty} \left( \frac{\mathcal E_1}{\mathcal E_0 -\epsilon }
+ \frac{\mathcal E_0 - \epsilon}{q-\epsilon} \right ) \mathcal R_{op}.
\label{eq:lambda_op}
\end{equation}
Plugging the above expression for $\lambda_{op}$ into Eq.~(\ref{eq:dpdt_R}), we get
\begin{equation}
{A_s^{-1/2}}\mathcal R_{op} = \frac{\eta_{max} }{1+\frac{\tau_{op}}{2}} = \frac{\eta_{max} }{1-\ln \mathcal R_{op}} \, ,
\label{eq:R_op}
\end{equation}
which yields
\begin{equation}
e^{-\tau_{op}} \approx  A_{s} \, \left(\frac{\eta_{max}}{\ln \eta_{max}}\right)^2
\frac{1}{\left[1+f(-\ln \eta_{max})\right]^2} \ ,
\label{eq:tau_op}
\end{equation}
where
\begin{equation}
f(x)=\frac{1}{x} \left[ \ln x + 1 -\frac{1}{2} \ln A_s + \frac{\ln x}{x} \right] \ .
\end{equation}
Finally, inserting Eq.~(\ref{eq:tau_op}) into Eq.~(\ref{eq:expand1}), it is straightforward to find an approximate expression for $\eta_{op}$:
\begin{equation}
\eta_{op} \approx \left (1- \frac{1}{|\ln \eta_{max}|\left[1+f(-\ln \eta_{max})\right]} \right ) \eta_{max} .
\end{equation}
Therefore, in the limit of $\eta_{max} \to 0$, one can see $\eta_{op} \approx \eta_{max}$, not $\frac{1}{2} \eta_{max}$, and that the next correction is logarithmic and therefore quite slow. This calculation confirms that the linear irreversible thermodynamics slightly out of equilibrium in~\cite{eff_Broeck} should \textit{not} be applicable in our case, simply because our processes are far from equilibrium. Indeed, in our case, the entropy production is maximal in the limit of $\eta_{max} \to 0$. In future studies, the validity of $\eta_{op} \approx \eta_{max}$ should be addressed in the context of universality for general information engines showing the maximum efficiency at the same point where the entropy production is maximal.

\section{Conclusions}

In this work, we have studied a simple example of an information engine which can be realized physically in terms of stochastic Markov processes. In agreement with previous studies, we find that the information feedback allows one to extract work in a situation where this would be thermodynamically impossible without feedback. Moreover, we confirm that total entropy production during relaxation obeys a fluctuation theorem, implying that the extracted work is bounded \textit{from above} by the mutual information gain between memory and system.

Providing a physical realization of the memory and the feedback loop, we have shown that the measurement process exhibits similar properties which are opposite in character. In particular, the entropy production during measurement is found to obey a fluctuation theorem as well. This implies that the measurement process itself costs energy, and that this additional energy supply is bounded \textit{from below} by the same mutual information gain. Putting these pieces together, it is no surprise that the total setup consisting of system and memory satisfies the conventional second law of thermodynamics. Thus, we have shown that the thermodynamic second law, which is required to hold for the entire system during any finite process, leads to a duality in the properties of system and memory in this kind of information engines.

For simplicity, we have presented most of our analytic results in the limit of infinite measurement- and relaxation time. However, the extension to finite times is straightforward. At the end of the paper, we have explicitly described some numerical results for finite-time measurement and relaxation. As in conventional heat engines the efficiency of the information engine is maximized when the cycle time becomes infinite. However, in contrast to conventional heat engines, the entropy production is also maximal in this limit. On the other hand, we have demonstrated that the power gain acquires its maximum at a finite cycle time. We have also discussed the relation between the maximal efficiency and the efficiency at the operating point of maximal power.

The striking differences between our model and conventional reversible heat engines can be traced back to the fact that our setup operates under non-equilibrium conditions. It would be interesting to investigate to what extent our observations can be explained in a universal framework.

\noindent\textbf{Acknowledgments}

\noindent
This research was supported by the NRF Grant No.2013R1A6A3A03028463 (J.U.), 2013R1A1A2011079 (C.K.), and  2013R1A1A2A10009722 (H.P.).
We thank the Galileo Galilei Institute for Theoretical Physics for the hospitality and the INFN for partial support during the completion of this work.


\end{document}